\documentclass{iopart}
\usepackage{epsfig}
\newcommand{\bee}{\begin{equation}}
\newcommand{\eeq}{\end{equation}}

\def\b{\begin{equation}}
\def\e{\end{equation}}
\def\br{\begin{eqnarray}}
\def\er{\end{eqnarray}}
\def\pa{\partial}
\def\l{\left}
\def\r{\right}

\def\d{{\mathrm d}}

\def\d{\displaystyle}

\def\ss{\scriptscriptstyle} 
\def\pl{Painlev\'{e} }

\begin{document}
\jl{6}
\title{Hawking radiation in different coordinate settings: Complex paths 
approach}
\author{S.~Shankaranarayanan, T.~Padmanabhan and K.~Srinivasan}
\address{IUCAA, Post Bag 4, Ganeshkhind, Pune 411 007, INDIA.} 
\ead{shanki@iucaa.ernet.in,paddy@iucaa.ernet.in}
\date{}
\begin{abstract}
We apply the technique of complex paths to obtain Hawking radiation in
different coordinate representations of the Schwarzschild
space-time. The coordinate representations we consider do not possess
a singularity at the horizon unlike the standard Schwarzschild
coordinate.  However, the event horizon manifests itself as a
singularity in the expression for the semiclassical action. This
singularity is regularized by using the method of complex paths and we
find that Hawking radiation is recovered in these coordinates
indicating the covariance of Hawking radiation as far as these
coordinates are concerned.
\end{abstract}

\section{Introduction}
       Quantum field theory(QFT) requires the notion of a time-like
killing vector field to define particles. In Minkowski space-time
existence of a global time-like killing vector, invariant under Lorentz
transformation, allows us to identify a preferred vacuum state for the
theory. Even in flat space-time, when curvilinear coordinate
transformations are allowed there arises other possible killing vector
fields which are time-like in part of the manifold.  Particle states --- in
particular vacuum state --- can be defined using these killing vector
fields. In general, these definitions will not be equivalent and the
Minkowski vacuum will appear to be a many particle state according to the
new definition. For example, the vacuum state of the Minkowski observer
appears to be a thermal state in the uniformly accelerated frame with a
temperature given by $T = g/2 \pi$ ($g$ is the constant acceleration).
\par
In a general curved space-time, we may not have any time-like killing
vector field and hence it is not always possible to identify a
preferred set of positive frequency modes. In static space-times,
which have a time-like killing vector, it is possible to define
positive frequency modes. However in these space-times, there could
arise more than one time-like killing vector field making the
quantization in different coordinates (describing the same
gravitational background) inequivalent. Hence, the concept of a
particle is not generally covariant in curved space-times.
\par  		
A spectacular prediction of quantum field theory in curved space-time
is the Hawking radiation from a black hole. Hawking\cite{hawking75}
showed that QFT in the background of a body collapsing to a black
hole, of mass $M$, will lead, at late times, to a radiation of
particles in all modes of the quantum field, with characteristic
thermal spectrum at a temperature equal to $(1/8\pi M)$. It is
generally believed that the pair production occurring near the horizon
in the case of black holes is ``real" with the mass of the black hole
being converted into energy of the emitted particles.
\par
But as mentioned in the earlier paragraphs, the concept of a
particle in quantum field theory is not generally covariant and
depends on the coordinates chosen to describe the particular
space-time.  It is therefore of interest to study the Hawking effect
in other coordinate representations.  The original method used by
Hawking is closely tied to the standard Schwarzschild
coordinate. Hawking's calculation of the particle creation amplitudes
(and Bogolubov coefficients) requires the knowledge of the wave modes
of the quantum field in the standard Schwarzschild coordinate. The
solutions of the wave equation in the Schwarzschild coordinate system 
cannot be written down in terms of simple functions and hence, the 
calculation of Bogolubov coefficients to identify the spectrum of 
radiation becomes intractable. Hence, it is necessary to have a method 
which does not use wave modes to calculate the emission spectrum.
\par
Hartle and Hawking~\cite{hawking76} obtained particle production in
the standard black hole space-times using a semiclassical analysis
which does not require the knowledge of the wave modes. In this
method, the semiclassical propagator for a scalar field propagating in
the maximally extended Kruskal manifold is analytically continued in
the time variable $t$ to complex values.  This analytic continuation
gives the result that the probability of emission of particles from
the past horizon is not the same as the probability of absorption into
the future horizon. The ratio between these probabilities is of the
form
\b
P[{\textrm{emission}}] = P[{\textrm{absorption}}] e^{-\beta E}
\label{eqn:thermalsys},
\e
where $E$ is the energy of the particles and $\beta^{-1}=(1/8\pi M)$
is the standard Hawking temperature. The above relation is interpreted
to be equivalent to a thermal distribution of particles in analogy
with that observed in any system interacting with black body
radiation. In this analysis, the probability amplitude for the
emission/absorption is calculated by identifying a particular path for
both these processes in the fully extended Kruskal
manifold. Unfortunately, the Kruskal extension plays a vital role in
obtaining the thermal spectrum in this analysis, and hence, it cannot
be adapted to other coordinate systems.

In Ref.~\cite{kt99}, the authors have obtained Hawking radiation {\it
without using the Kruskal extension}. They have shown that the
coordinate singularity present at the horizon, in the standard
Schwarzschild coordinate, manifests itself as a singularity in the
expression for the classical action $S_0$ which occurs in the
semiclassical propagator $K(r_2, t_2; r_1, t_1)$, which is given by
\b
K(r_2, t_2; r_1, t_1) = N \exp\l(\frac{i}{\hbar}S_0(r_2, t_2; r_1, t_1) \r).
\e
\noindent $S_0$ is the action functional satisfying the classical
Hamilton Jacobi (HJ) equation for a massless particle to propagate
from $(t_1, r_1)$ to $(t_2, r_2)$ and $N$ is the suitable
normalization constant. The authors used the method of complex paths
discussed in Ref.~\cite{landau3} where it is used to describe
tunneling processes in non-relativistic semi-classical quantum
mechanics. The method of complex paths used in ordinary quantum
mechanics was modified appropriately to produce a prescription that
regularizes the singularity in the action functional and Hawking
radiation was recovered as a consequence. This method also provides a
heuristic picture of Hawking radiation as tunneling. In this method,
by construction, the {\it in} modes of the field is in the vacuum
state and the tunneling process near the horizon produces a steady
out-flux of radiation.  The authors consider Unruh vacuum to define
particles at future infinity.  (See for example,
Refs.~\cite{candelas,visser}).

In this work, we apply the method of complex paths to two non-static
coordinate systems of the Schwarzschild space-time and obtain Hawking
radiation. The two coordinate systems we consider are: Lemaitre
coordinate (a time dependent system) and \pl coordinate (a stationary
--- but not a static --- system). The Lemaitre coordinate being a time
dependent system, suggests that there could be a genuine particle
production. Hence, it is interesting to know what kind of particle
spectrum is seen in this coordinate system. In the case of \pl
coordinate, a priori it not  clear whether there exists a genuine
particle production!

Also there has been a growing interest in the semiclassical treatment
of Hawking radiation in the \pl metric (see Refs.~\cite{parikh,ralf})
in relation with the hydrodynamic analog models of event horizon. The
hydrodynamic analog of an event horizon was suggested by
Unruh~\cite{unruh81} as a more accessible phenomenon which might shed
some light on the Hawking effect and, in particular, on the role of
ultra-high frequencies~\cite{jacobson}. An event horizon for the sound
waves appears in principle wherever there is a closed surface through
which a fluid flows inwards at the speed of sound, the flow being
subsonic on one side of the surface and supersonic on the other. There
is a close analogy between sound propagation on a background
hydrodynamic flow, and field propagation on a background in a curved
space-time. Determining whether and how sonic black holes radiate
sound, in a full calculation beyond the hydrodynamic approximation or
in an actual experiment, can thus offer some suggestions about
black-hole radiance and its sensitivity to ultra-high frequency
physics. It is seen that for an irrotational fluid with constant
density and sound speed with the fluid velocity ${\bf v} = \sqrt{2 G
m/r}$, then one can obtain a metric element which is conformal to \pl
form of the Schwarzschild space-time. The acoustic horizon of this
kind are of interest in understanding the role of ultra-high
frequencies in the Hawking radiation.  

In these two coordinate representations, \pl and Lemaitre, metric has
no coordinate singularity at the horizon. We notice that the action
function for a classical particle in both these coordinates acquires a
singularity at the horizon. In Ref.~\cite{kt99}, it was shown that it
is only the singularities that appear in the action that contribute to
particle production. For the Lemaitre coordinate, the regularization
of the singularity in action is achieved and the Hawking radiation is
recovered without reference to the time dependent nature of the
system. The same feature is seen even in the stationary system where
the singularity represented by the horizon appears as a pole in the
expression for the action and upon regularization by a suitable
prescription gives Hawking radiation. The temperature associated with
the radiation for both the coordinate representations is same as that
obtained in the standard Schwarzschild coordinates. We would like to
note the following: Even-though the action satisfies a generally
covariant HJ equation, when one transforms coordinates, the action
integral develops poles and this need to be interpreted. In the case
of Kruskal coordinate, which is the maximal extension of Schwarzschild
space-time, it is easy to show that the semiclassical action when
expressed in terms of Kruskal coordinates does not contain the
singularity. [The HJ equation (of a massless particle) when expressed
in terms of the Kruskal coordinates $(V, U,\theta, \phi)$ is of the
form $\left( {\partial S_0/\partial V} \right)^2 - \left( \partial
S_0/\partial U \right)^2 = 0$ (for S-wave i.e $l = 0$).  The solution
of the equation can be easily obtained and is given by $ S_0(V_2,U_2;
V_1,U_1)= S_0(2,1) = -p_V (V_2-V_1) \pm p_U (U_2 - U_1) $.]

To our knowledge, the semi-classical treatment of Hawking radiation
for the Lemaitre coordinate has not been performed in the
literature. The semi-classical treatment of Hawking radiation for the
\pl metric is obtained recently by Parikh and
Wilczek~\cite{parikh}. The authors considered Hawking radiation as a
pair creation outside the horizon, with the negative energy particle
tunneling into the black hole. They have not considered the
contribution of the tunneling of a particle from inside the horizon to
outside the horizon to describe the Hawking radiation. Our approach is
quite different and could be of more general validity.

Before proceeding to the technical aspects, it is necessary to outline
certain conceptual issues regarding the emission spectrum from the
black hole in different coordinate systems. The wave modes obtained
using semiclassical techniques, in general, are the exact modes of the
quantum system in the asymptotic regions. Thus, if the asymptotic
structure of the space-time is different for any two coordinates, then
the semiclassical wave modes associated with these two coordinate
systems will themselves be different. Hence, the spectrum of the
radiation emitted, if any, in these two coordinates, with different
asymptotic structure, will be completely different. The Lemaitre
coordinate can be modeled as that natural to a freely falling observer
whose velocity at radial infinity is zero [for more discussion see
Sec.~(\ref{subsec:lemaitre})]. In the asymptotic past and future, the 
structure of these two coordinate systems are similar;  suggesting 
that the asymptotic wave modes must be the same. In Sec.~(\ref{sec:semiclas}),
we have shown that the spectrum of the particles emitted in the two
coordinates, \pl and Lemaitre, is a thermal spectrum --- same as in
standard Schwarzschild.

Even though, the two coordinate systems, \pl and Lemaitre, possess no
singularities at the horizon, the action function for a classical
particle acquires singularity at the horizon. The results of
Ref.~\cite{kt99}  show that the contribution to the thermal
radiation of the emitted particles is from the singularities. Thus, it
suggests that the temperature associated with the thermal spectrum in
these two coordinate systems, \pl and Lemaitre, should be the same as
that of the standard Hawking temperature.  The metric corresponding to
the standard Schwarzschild coordinate is time reversal invariant, $t
\rightarrow -t$. However, the metric corresponding to the two
coordinates, \pl and Lemaitre, are not time reversal invariant.  The
time reversal transformation in these cases correspond to a different
physical situation. However, the line elements corresponding to these
different physical situations are related to the standard
Schwarzschild by a coordinate transformation [see
Sec.~(\ref{sec:coor}) for more details].

The space-time structure of these two coordinates, Lemaitre and \pl,
are best analyzed using R and T regions introduced by
Novikov~\cite{novikov64}. The analysis of the space-time structure of
the two coordinates, \pl and Lemaitre, using R and T regions provides
an elegant method in understanding the global structure of the
space-time, thus making the detailed analysis of particle trajectories
unnecessary. Using this analysis, we show that the part/whole of the
Lemaitre/\pl manifold is doubly mapped {\it w.r.t} the standard
Schwarzschild space-time. We find that the complex paths we use in our
analysis, takes into account of all possible paths satisfying the
semiclassical ansatz -- irrespective of he multiple mapping of the
part/whole of the space-time. The above statement can be better
appreciated in the language of path-integrals.

The Feynman propagator $G_F(x,y|g)$ for a scalar field, of mass $m$,
propagating in a background metric $g_{ik}$ can be written in two
equivalent forms as 
\br G_F(x,y|g)& =& \sum_{paths} \exp(-m
{\mathcal{R}}(x,y)) \\ &=& \int_0^{\infty} d\tau \exp(-m^2 \tau) \int
{\mathcal{D}}x \exp\l(-\frac{1}{4} \int_0^\tau g_{ik} \dot{x}^i
\dot{x}^k d\eta \r) \nonumber \er where ${\mathcal{D}}x$ is the
measure of the path. In the first form, ${\mathcal{R}}(x,y)$ is the
proper length of a path connecting the events $x$ and $y$, calculated
with the background metric $g_{ik}$ and the sum is over all paths. It
is also possible to show that the result is equivalent to the second
expression. In the case of space-times where there are no multiple
mappings then the measure (of all sets) of paths will be the
same. However, in the space-times where part/whole of the manifold is
multiply mapped, the measure of certain sets of paths will be
different from certain other sets of paths. Hence, it is necessary to
do a global analysis of the background space-time to identify the
measure of the paths. Similarly, in the method of complex paths, the
family of paths we use to calculate the emission and absorption
probability takes into account of all paths irrespective of the
multiple mapping of the part/whole of the space-time. These facts will
have bearing on the interpretation of temperature as explained in
Sec.~(\ref{sec:semiclas}).  

Our result is of physical significance for the following reason: it is
normally assumed that the evaporation process results from an
instability of the vacuum in the presence of background metric. The
particles are produced at a constant rate suggesting that the Hawking
radiation converts the mass of the black hole into energy, thereby
decreasing the mass. The decrease in the black hole mass is a physical
effect and should be independent of the coordinate transformations and
hence {\it Hawking radiation should be covariant}. Here we show that
this is indeed true by the method of complex paths as far as these
coordinates are concerned.

Before concluding this section, we would like to stress the following
point: There is no correspondence between the particles detected by a
detector (which is coupled to a quantum field $\phi$) and the particle
spectrum obtained by the field theoretic analysis. This result is
known in the context of flat space-times. For example, in the case of
uniformly rotating detector, the expectation value of the rotational
number operator in the Minkowski space-time is zero whereas the
response of a uniformly rotating Unruh-Dewitt detector is non-zero
(For more examples see,
Refs.~\cite{schrambo98,letaw,paddycqg,bandd82}). In the case of
classical gravitational backgrounds, defining particles using the
particle detectors is non-trivial. The trouble in defining the
particle detectors is that the inertial detectors become free falling
detectors and in general no two free falling observers will agree on a
choice of vacuum. Only in exceptional cases in high symmetry will a
set of detector exist that all register no particles, and this set may
not even be free falling. These facts will have important implications
in comparing the results of a freely falling detector in the
Schwarzschild space-time \cite{davies76} to that of the particle
spectrum we obtain for the Lemaitre coordinates in
Sec.~(\ref{subsec:lemaitres}).

The organization of the paper is as follows. In
Sec.~(\ref{sec:hawking}), the method of complex paths, as applied to
Schwarzschild like space-times, is briefly recapitulated. In
Sec.~(\ref{sec:coor}), we briefly describe the definition of R and T
regions of the spherically symmetric space-times and also discuss the
properties of the two coordinates which is of our interest, namely
Lemaitre coordinate and \pl metric. In Sec.~(\ref{sec:semiclas}), we
apply the method of complex paths to the two coordinate systems and
obtain the temperature corresponding to the radiation. Finally in
Sec.~(\ref{sec:conclusions}) we summarize the results of this paper.

Through out this paper, the metric signature we shall adopt is $(+ - -
-)$ and the quantum field is a scalar field.

\section{Hawking radiation in Schwarzschild coordinate}
\label{sec:hawking}
In this section, we briefly describe the recovery of Hawking radiation 
in the Schwarzschild coordinates using the method of complex paths.  
We do this in $(1+1)$ dimensions since all the physics is contained in the 
$(t,r)$ plane. The generalization to a massive field and to $(1+3)$ dimensions 
is straight forward. Consider a certain patch of space-time in $(1+1)$ 
dimensions, which -- in a suitable coordinate system -- has the line element, 
(with $c=1$) 
\begin{equation}
ds^2 = B(r)dt^2 - B^{-1}(r) dr^2  \label{eqn:metric1}
\end{equation}
where $B(r)$ is an arbitrary function of $r$.   
We assume that the function $B(r)$ vanishes at some $r = r_0$ 
with $B'(r)=dB/dr$ being finite and nonzero at $r_0$. 
The point $r=r_0$ indicates the presence of a horizon.
It can be easily verified that no physical singularity exists 
at the horizon since the curvature invariants do not have a 
singularity on the horizon. Therefore, near the horizon, we expand $B(r)$ as
\begin{equation}
B(r) = B'(r_0)(r - r_0) + {\cal O}[(r-r_0)^2] = R(r_0)(r-r_0).
\label{eqn:horizon}
\end{equation}
\noindent where it is assumed that $R(r_0) \neq 0$. 
\par
The equation satisfied by the minimally coupled massless scalar field in the 
above background metric is, 
\begin{equation}
{1 \over B(r)} {\partial^2 \Phi \over \partial t^2} -
{\partial \over \partial r} \left( B(r) {\partial \Phi 
\over \partial r} \right) = 0 .
\label{eqn:kgr}
\end{equation}
The semiclassical wave functions satisfying the above equation 
are obtained by making the standard ansatz for $\Phi$ which is,
\begin{equation}
\Phi(r,t) = \exp\left[ {i \over \hbar} S(r,t)\right] 
\label{eqn:semi}
\end{equation}
where $S$ is a function which is expanded in powers of $\hbar$ of the form
\begin{equation}
S(r,t) = S_0(r,t) + \hbar S_1(r,t) + {\hbar}^2 S_2(r,t) \ldots 
\label{eqn:exp}.
\end{equation}
Substituting into the wave equation~(\ref{eqn:kgr}) and neglecting 
terms of order $\hbar$ and greater, we find to the lowest 
order,
\begin{equation}
{1 \over B(r)} \left( {\partial S_0 \over \partial t} 
\right)^2 - B(r)\left( {\partial S_0 \over \partial r} 
\right)^2 = 0 
\label{eqn:seqn2}.
\end{equation}
Eq.~(\ref{eqn:seqn2}) is just the Hamilton-Jacobi equation 
satisfied by a massless particle moving in the space-time 
determined by the line element~(\ref{eqn:metric1}).  
The solution to the above equation is 
\begin{equation}
S_0(r_2,t_2; r_1,t_1)= S_0(2,1) = -E(t_2-t_1) \pm E 
\int^{r_2}_{r_1} {dr \over B(r)}.  
\label{eqn:ssol3}
\end{equation}
The semiclassical kernel $K(r_2,t_2; r_1,t_1)$ for 
the particle to propagate 
from $(t_1,r_1)$ to $(t_2,r_2)$ in the saddle point 
approximation can be written down immediately as.
\begin{equation}
K(r_2,t_2; r_1,t_1) = N \exp\left( {i\over \hbar} 
S_0(r_2,t_2; r_1,t_1) \right),
\end{equation} 
where $S_0$ is the action functional satisfying the classical 
Hamilton-Jacobi equation in the 
massless limit and $N$ is a suitable normalization constant.  

The sign ambiguity (of the square root) in Eq.~(\ref{eqn:ssol3}) is 
related to the ``outgoing'' ($p_r = \partial S_0 /\partial r\,>0$) or 
``in-going'' ($\partial S_0 /\partial r\,<0$) nature of the particle. 
[The momentum of the particle moving in $r$ is given by 
$p_r = \partial S_0/\partial r $. Hence, $p_r > 0$ corresponds to 
outgoing particle {\it w.r.t} the horizon.] As long as points $1$ 
and $2$, between which the transition amplitude is calculated, are 
on the same side of the horizon (i.e. both are in the region 
$r>r_0$ or in the region $r<r_0$), the integral in the action is 
well defined and real. But if the points are located on opposite 
sides of the horizon then the integral does not exist due to 
the divergence of $B^{-1}(r)$ at $r=r_0$.

Therefore, in order to obtain the probability amplitude for 
crossing the horizon we have to give an extra prescription 
for evaluating the integral \cite{paddy91}.  
Since the horizon defined by $B(r_0)=0$ is null we may carry out 
the calculation in Euclidean space or ---equivalently---use 
an appropriate $i\epsilon$ prescription to specify the 
complex contour over which the integral has to be performed 
around $r=r_0$.  
The prescription we use is that we should take the contour 
for defining the integral to be an infinitesimal semi-circle 
{\it above} the pole at $r=r_0$ for outgoing particles on 
the left of the horizon and in-going particles on the right.  
Similarly, for in-going particles on the left and outgoing 
particles on the right of the horizon (which corresponds to 
a time reversed situation of the previous cases) the contour 
should be an an infinitesimal semi-circle {\it below} the 
pole at $r=r_0$. 
Equivalently, this amounts to pushing the singularity at 
$r=r_0$ to $r = r_0 \mp i\epsilon$ where the upper sign 
should be chosen for outgoing particles on the left and 
in-going particles on the right while the lower sign should 
be chosen for in-going particles on the left and outgoing 
particles on the right. 
For the Schwarzschild case, this amounts to adding an 
imaginary part to the mass since $r_0 = 2M$. 
\par
The prescription outlined above has its origin and basis in the  
method of complex paths which is  outlined in Ref.~\cite{kt99} 
(see also~\cite{landau3}). This method is used to compute the 
transmission and reflection coefficients in standard semiclassical 
quantum mechanics by specifying a suitable complex contour for a given 
tunneling scenario. This contour is chosen between two semiclassical 
regions (where the wave function can be approximated using the 
semiclassical ansatz with negligible error) such that the semiclassical 
approximation holds all along the contour. If singularities, which 
represent distinctive features of the system under consideration, are 
present in the quantum system and these lie between the semiclassical 
regions, the appropriate complex contour contains useful information 
that decides the steady state behavior of the system. The complex path 
is identified from a region, say $L$ (a superposition of the incident and 
the reflected wave), to a second region $R$ (just contains the transmitted 
wave) is identified from $R$ to $L$ such that (a) all along the path the 
semiclassical ansatz is valid and (b) the reflected wave is exponentially 
greater than the incident wave atleast in the latter part of tha path near 
the region $L$. The transmitted wave is then moved along the path to obtain 
the reflected wave and thus the amplitude of reflection is identified in 
terms of the transmission amplitude. Having done this, the normalization
condition is used, {\it i.e.}, the sum of the modulus square of the
transmission and reflection amplitudes should equal unity, to determine the
exact values of the transmission and reflection coefficients. 
\par
In the black hole space-times considered in this section, 
the singularity that appears in the action functional in 
Eq.~(\ref{eqn:ssol3}) is directly attributable to the presence 
of a horizon. Since the semiclassical approximation is applicable on 
either side of the horizon and arbitrarily close to it, 
the complex contours needed to bypass the singularity 
follow from the demand that the semiclassical approximation 
hold all along the contour. The type of singularity encountered 
here is similar to that encountered in the one-dimensional 
Schr\"odinger system with a potential of the form $(-1/x^2)$. The 
method of complex paths gives the appropriate contours when dealing 
with right moving or left moving waves propagating across the 
singularity at $x=0$. 
 
Consider therefore, an outgoing particle 
($\partial S_0 /\partial r\,>0$) at $r=r_1<r_0$. 
The modulus square of the amplitude for this particle to 
cross the horizon gives the probability of emission of 
the particle.  
The contribution to $S_0$ in the ranges $(r_1,r_0-\epsilon)$ 
and $(r_0+\epsilon,r_2)$ is real. 
Therefore, choosing the contour to lie in the upper 
complex plane,
\begin{eqnarray}
S_0[{\rm emission}] &=& -E\lim_{\epsilon \to 0} 
\int^{r_0+\epsilon}_{r_0-\epsilon} {dr \over B(r)} 
+ \; ({\rm real \; part})  \nonumber \\
&=& {i \pi E \over R(r_0)}+ ({\rm real \; part})
\end{eqnarray}
where the minus sign in front of the integral corresponds 
to the initial condition that $\partial S_0 /\partial r\,>0$ 
at $r=r_1<r_0$.  
For the sake of definiteness we have assumed $R(r_0)$ in 
Eq.~(\ref{eqn:horizon}) to be positive, so that 
$B(r)<0$ when $r<r_0$.  
(For the case when $R<0$, the answer has to be modified by a 
sign change.)  
The same result is obtained when an in-going particle 
($\partial S_0 /\partial r\,<0$) is considered at 
$r=r_1<r_0$.  
The contour for this case must be chosen to lie in the 
lower complex plane. 
The amplitude for this particle to  cross the horizon  is 
the same as that of the outgoing particle due to the time 
reversal invariance symmetry obeyed by the system.     

For an in-going particle ($\partial S_0 /\partial r\,<0$) at $r=r_2>r_0$, 
choosing the contour to lie in the upper complex plane, we get, 
\b
S_0[{\rm absorption}] = -{i \pi E \over R(r_0)} + ({\rm real \; part})
\e
The same result is obtained when an outgoing particle 
($\partial S_0 /\partial r\,>0$) is considered at $r=r_2>r_0$.  
The contour for this case should be in the lower complex 
plane and the amplitude for this particle to cross the 
horizon  is the same as that of the in-going particle due 
to time reversal invariance.

Taking the modulus square to obtain the probability $P$, 
we get,
\begin{equation}
P[{\rm emission}] = \exp\left(-{4 \pi E \over \hbar R} 
\right)P[{\rm absorption}]. \label{eqn:ssys}
\end{equation}
Using the interpretation of Hartle and Hawking \cite{hawking76}, the above
relation  shows the thermal behavior of emitted particles. Comparing 
Eq.~(\ref{eqn:thermalsys}) and Eq.~(\ref{eqn:ssys}), we identify the 
temperature of the horizon in terms of $R(r_0)$. Eq.~(\ref{eqn:ssys}) is based 
on the assumption that $R>0$. If $R<0$, one can show that, there will be a 
change of sign in the equation. Incorporating both the cases, the general 
formula for the horizon temperature $\beta^{-1} = \hbar |R|/4\pi$.

For the Schwarzschild black hole, $R = (2M)^{-1}$, and the temperature is  
$\beta^{-1}= \hbar /8\pi M$. For the de Sitter space-time, $R = H$, giving 
$\beta^{-1}= \hbar H/2\pi $. Although Reissner-Nordstrom space-time has a 
complicated space-time structure, using the method of complex paths we can 
obtain the correct temperature in a straight forward manner. The prescription 
given for handling the singularity is analogous to the analytic continuation 
in time proposed by Hartle and Hawking~\cite{hawking76} to derive Black hole 
radiance.  

\section{Non-singular coordinates in Schwarzschild manifold}
\label{sec:coor}
In this section, we discuss the properties of the two coordinates, \pl 
and Lemaitre, which are of our interest in this paper. The space-time 
structure of these two coordinates are best analyzed using R and T regions 
introduced by Novikov~\cite{novikov64}. For the sake of completeness, we 
briefly outline key features of R and T regions of spherically symmetric 
space-times.
 
\subsection{R and T regions for spherically symmetric space-times}
\label{subsec:rtregions}
The general expression for the interval in a spherically symmetric field 
can be written in the following 
form:
\b
ds^2 = e^{\nu} (dx^0)^2 - e^{\lambda} (dx^1)^2 
- e^{\mu} d\Omega^2.
\label{sphesymm}
\e
\noindent Here $x^0$ is the time coordinate, $x^1$ is the radial space 
coordinate, and $d\Omega^2$ is the angular line element. [$\nu$,
$\lambda$ and $\mu$ are the functions of $x^0$ and $x^1$.] The reason
for such a general line element for centrally symmetric gravitational
field because the choice of ``radius vector" is arbitrary. This is due
to the fact that there is no quantity which has all the properties of
the Euclidean radius vector. Hence, because of this arbitrariness in
the choice of reference in general relativity, one can subject the
coordinates to any transformation which does not destroy the spherical
symmetry of the metric.
\par
We assume that, at spatial infinity, the metric is Euclidean, and the 
expression for the interval in spherical coordinates becomes
\b
ds^2|_{\infty} = (dx^0)^2 - (dx^1)^2 - (x^1)^2 d\Omega^2.
\e
\noindent Changes of the coordinates $x^0$ and $x^1$ 
\b
x^0 = x^0(x^{0'}, x^{1'}),~~~~ x^1 = x^1(x^{0'}, x^{1'})
\label{trans1} 
\e
\noindent leave the spherical symmetry unaffected. Using a transformation of 
this type, we set
\b
x^1 =  e^{\mu(x^{0'}, x^{1'})/2}
\label{trans2}
\e
\noindent where the old coordinates have primes; then we select $x^0 = 
x^0(x^{0'},x^{1'})$ so that the metric component $g_{10}$ vanishes, which is 
always possible. The expression for the interval in Eq.~(\ref{sphesymm}) 
takes on the form
\b
ds^2 =  e^{\nu(x^0, x^1)} (dx^0)^2 - e^{\lambda(x^0, x^1)} (dx^1)^2 
- (x^1)^2 d\Omega^2.
\label{modspex}
\e
The transformation in (\ref{trans2}), which brings the interval to the
form in (\ref{modspex}) can, however, lead to a singular metric. The
transformation in (\ref{trans2}) can also lead to line elements of the
form in (\ref{modspex}), in which the metric component $g_{00}$ is
negative and $g_{11}$ is positive. In such cases, the quantity
$(g_{11})^{1/2}dx^1$ measures the proper time of a particle in this
coordinate system implying that $x^1$ should be treated as the time
coordinate. Hence, the expression for the interval will read
\b
ds^2 = e^{\nu(x^0, x^1)} (dx^0)^2 - e^{\lambda(x^0, x^1)} (dx^1)^2 
- (x^0)^2 d\Omega^2.
\label{modspet}
\e
\noindent Thus, by using a transformation of the type given in 
(\ref{trans2}), one can bring the interval of a spherical
gravitational field either to the form in (\ref{modspex}) or to the
form in (\ref{modspet}). If at the given event in the general
coordinate system in (\ref{sphesymm}), the inequality,
\b
e^{\nu - \lambda} > \l( \frac{\pa\mu}{\pa x^0}/ \frac{\pa\mu}{\pa x^1}\r)^2,
\label{maincond}
\e
\noindent is satisfied, then the event is in an R-region. If the opposite 
inequality is satisfied, the event is in a T-region. The equality
condition is an additional condition to the system of the
gravitational field equations and therefore for an arbitrary
spherically symmetric distribution in vacuum, it may not be possible
to fulfill it in any four-dimensional region. The equality determines
the boundary between R- and T- regions.
\par
	 The definitions of R- and T- regions can be shown to be coordinate 
invariant. These will be found useful in understanding global space-time 
structure of Lemaitre and \pl coordinates.

\subsection{The Lemaitre coordinate system}
\label{subsec:lemaitre}
We consider here the Lemaitre coordinate system which is a set of non-static 
coordinates like the Kruskal coordinate. The coordinate transformation from 
the Schwarzschild coordinates $(t,r)$ to Lemaitre coordinates are,  
\br
\!\!\!\!\!\!\!\!\!\!\!\!\!\!\!\!\!\!\!\!\!\!\!\!\!\!\!\!\!\!\!\!\!\!\!\!\!\tau& =& \pm t \pm \sqrt{2M} \int\! {\sqrt{r} dr\over r-2M} 
=\pm t \pm 2M\l\{\ln\l( { 1-\sqrt{r/2M} \over 1+ \sqrt{r/2M}}\r) 
+ 2\sqrt{r\over 2M} \r\}
\nonumber \\
\label{lemaitretrans}
\!\!\!\!\!\!\!\!\!\!\!\!\!\!\!\!\!\!\!\!\!\!\!\!\!\!\!\!\!\!\!\!\!\!\!\!\!R & = & t + {1 \over \sqrt{2M}} \int\! { r\sqrt{r}dr\over r-2M} =  
t + 2M \l\{ \ln \l[ { 1-\sqrt{r/2M} \over 1+ \sqrt{r/2M}}\r] 
+ 2\sqrt{r \over 2M} \l[ 1 + {r \over 6M}\r]\r\}
\er
Choosing the positive sign in the above transformations, the Schwarzschild 
line element becomes,
\b
\!\!\!\!\!\!\!\!\!\!\!\!\!\!\!\!\!\!\!\!\!\!\!\!\!\!\!ds^2 = d\tau^2 - \frac{dR^2}{\left[\frac{\d 3}{\d 4M}(R-\tau)\right]^{2/3}} 
-  (2M)^2\left[\frac{3}{4M}(R-\tau)\right]^{4/3}
(d\theta^2+\sin^2\theta d\phi^2).
\label{lemaitremetric1}
\e
It is easy to see from the transformations in Eq.~(\ref{lemaitretrans}) that 
the Schwarzschild $r$ coordinate is given by the relation
\b
\l({r \over 2M}\r) = \l[{3 \over 4M}(R-\tau)\r]^{2/3}
\e
Hence the horizon $r=2M$ corresponds to the value $3(R-\tau)/4M = 1$. 
Note that the quantity $(R-\tau)$ can never be negative since it
corresponds to the Schwarzschild radial coordinate. The coordinate $R$
is everywhere space-like while the coordinate $\tau$ is everywhere
time-like. The metric above is non-stationary and test
particles at rest relative to the reference system are particles
moving freely in the given field. It can be easily shown (see
Ref.~\cite{landau2}) that even in this coordinate, particles cannot
remain at rest for $r<2M$.
\par
If we choose minus sign in Eq.~(\ref{lemaitretrans}), the Schwarzschild 
line element becomes
\br
\!\!\!\!\!\!\!\!\!\!\!\!\!\!\!\!\!\!\!\!\!\!\!\!\!\!\!ds^2 = d\tau^2 - 
\frac{dR^2}{\left[\frac{\d 3}{\d 4M}(R+\tau)\right]^{2/3}} 
- (2M)^2\left[\frac{3}{4M}(R+\tau)\right]^{4/3}
(d\theta^2+\sin^2\theta d\phi^2).
\label{lemaitremetric2}
\er
Note that the sign of $\tau$ has changed. This metric describes a
system where particles trajectories move outward from the singularity
at $r=0$.  This is analogous to the part of Kruskal coordinate where
the past singularity and the time reversed Schwarzschild sector
represent such a feature. This metric will be used when the action for
outward moving particles needs to be considered. These coordinates can
be modeled as that natural to a freely falling observer whose velocity
at radial infinity is zero. The time coordinate $\tau$ measures the
proper time of free falling observers; each observer moves along a
line $ R = constant$. Figure (1) shows the Penrose diagram for the
Schwarzschild geometry. The line-element (\ref{lemaitremetric1})
covers $(R, T_-)$ and the line-element (\ref{lemaitremetric2}) covers 
$(R, T_+)$.
\begin{figure}[!htb]
\centerline{\epsfig{file=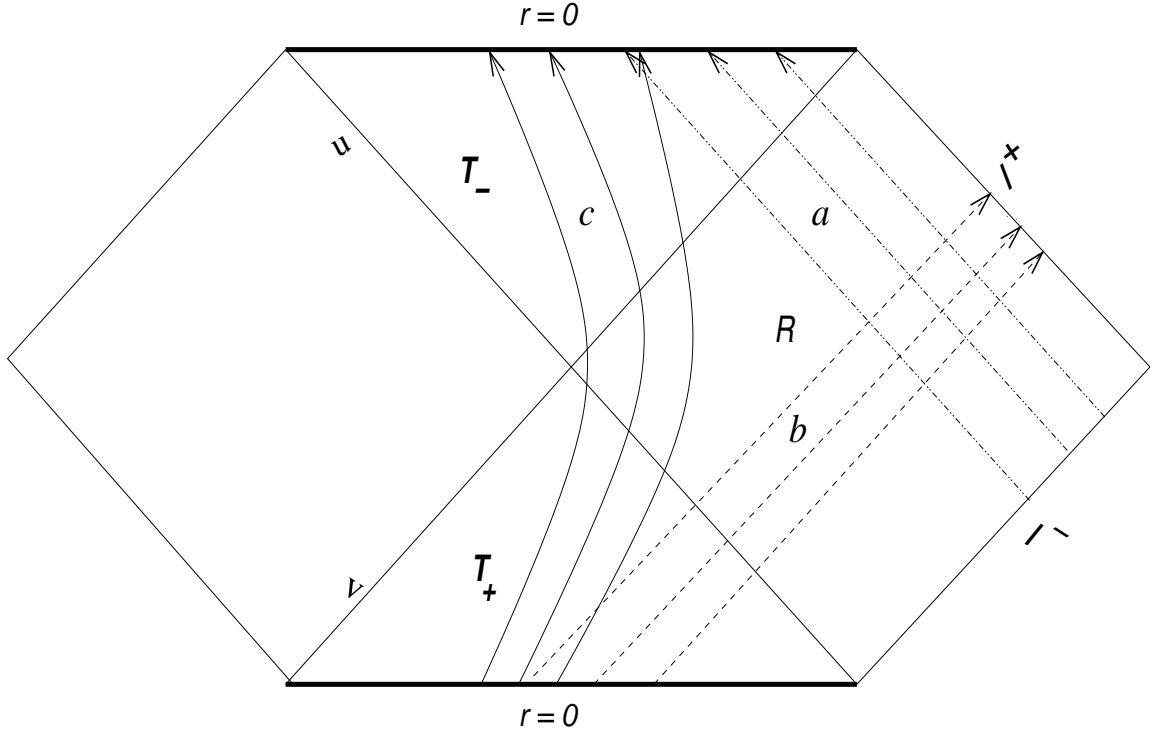,height=3.8in,width=6.0in}}
\label{F:lemaitre}
\caption{{\small The Penrose diagram of the Schwarzschild space-time. $u$ 
and $v$ are the lines $r = 2M$; they are the boundaries of the $R$ and
$T$ regions. $(R, T_-)$ region corresponds to the line-element
(\ref{lemaitremetric1}) while $(R, T_+)$ region corresponds to that of
(\ref{lemaitremetric2}). Lines $a$, $b$ and $c$ are the world lines of
freely falling particles. $a$ are the geodesics of particles falling
inward from $r = \infty$; $b$ are the geodesics of particles flying
outward form $r = 0$ to $r \to \infty$; and $c$ are the trajectories
of the particles which move outward which eventually falls into the
singularity.}}
\end{figure}
\par
Using the criteria for R- and T- regions, defined in 
Sec. (\ref{subsec:rtregions}), we find that the region outside the 
Schwarzschild sphere, for the line element in (\ref{lemaitremetric1}), i.e., 
where
\b
\frac{3}{2} (R - \tau) > 2M
\label{limcond}
\e
\noindent is a R-region; but the region inside the Schwarzschild sphere, 
where the inequality is opposite to the above inequality, is a T-region.
\par
In the T-region there is an obvious asymmetry in the direction of time flow. 
In the T-region (T$_{-}$), corresponding to the line element in 
(\ref{lemaitremetric1}), the test particles cannot remain at rest and these 
test particles will be directed towards $r = 0$ where the curvature invariants 
are infinite. The line element in (\ref{lemaitremetric2}) will define 
second type of T-regions --- an expanding region T$_{+}$, which has completely 
opposite properties. Here all bodies move away from the singularity to 
the R-regions. A vacuum T$_{+}$-region can be joined to the internal 
solution for an expanding sphere, but not to that for a contracting sphere 
as in the case for a T$_{-}$-region. For a physically realizable event, the 
coexistence of $T_+$ and $T_-$ regions is impossible. In other words, the T 
region of the standard Schwarzschild is a doubly mapped T region in Lemaitre.
\par
The above analysis will be useful for our semiclassical treatment
given in Sec. (\ref{subsec:lemaitres}). As mentioned earlier, the
complex paths we use in our analysis, takes into account of all
possible paths satisfying the semiclassical ansatz - irrespective of
the multiple mapping of the part of the space-time. In principle, if
the whole of space-time is doubly mapped {\it w.r.t.} the
Schwarzschild space-time, there will be two distinct R- and T-
regions. The complex paths, as used in Sec.~(\ref{sec:hawking}),
will have contributions from both of these which have no common point
due to each path being in different R- and T- regions. Hence, the
contribution to the amplitude of emission/absorption by these two
paths will be mutually exclusive. In the case of Lemaitre coordinate,
however, only part of the region(T-) is doubly mapped, it is always
possible to find one point that is common to the paths contributing to
absorption/emission. Hence, these paths are not mutually
exclusive. These paths, on the other hand, will be mutually exclusive
when one considers the probability amplitude -- which is the important
quantity in our approach. Hence, the action we obtain by regularizing
the singularity and the resulting probability amplitude, for
absorption/emission, will have equal contributions from both these
paths. 

\subsection{Painlev\'{e} metric}
\label{subsec:painleve}
In this section, we consider Painlev\'{e} coordinates, the other representation
 of the Schwarzschild space-time which is of our interest (for a discussion 
see Ref.~\cite{kraus}). The metric in these coordinates is non-singular at 
the horizon just as in the previous case but, unlike the Lemaitre metric, it 
is stationary. The coordinate transformation from the Schwarzschild 
coordinates $(t,r,\theta,\phi)$ to the Painlev\'e coordinates $(\tau_P,r,
\theta,\phi)$ is given by
\b 
\tau_P=t+2\sqrt{2M\,r}+2M\ln\l(\frac{\sqrt{r}-\sqrt{2M}}{\sqrt{r}+\sqrt{2M}}
\r)
\label{paint1}.
\e
Note that only the Schwarzschild time coordinate is transformed. The radial 
coordinate remains the same (as do the angular coordinates). 
The line element, in these new coordinates, is given by
{\small
\b
ds^2=\l(1-\frac{2M}{r}\r)d\tau_P^2-2\sqrt{\frac{2M}{r}}d\tau_P dr
- dr^2 - r^2 d\Omega^2
\label{painm1}.
\e
} 
The metric is stationary and the constant time slice is simply flat
Euclidean space. Another line element is easily obtained by
considering the transformation of the form
\b 
\tau_P=t-2\sqrt{2M\,r}-2M\ln\l(\frac{\sqrt{r}-\sqrt{2M}}{\sqrt{r}+\sqrt{2M}}
\r)
\label{paint2}.
\e
which leads to the equivalent line element
{\small
\b 
ds^2=\l(1-\frac{2M}{r}\r)d\tau_P^2+2\sqrt{\frac{2M}{r}}d\tau_P dr-dr^2-
r^2 d\Omega^2
\label{painm2}
\e
}
\noindent The line elements (\ref{painm1}) and (\ref{painm2}) can be 
written in a compact form 
\b
ds^2 = (d\tau_P)^2 - \l( dr \pm \sqrt{\frac{2M}{r}} d\tau_P \r)^2 - r^2
d\Omega^2. 
\label{painfull}
\e
\begin{figure}[!htb]
\centerline{\epsfig{file=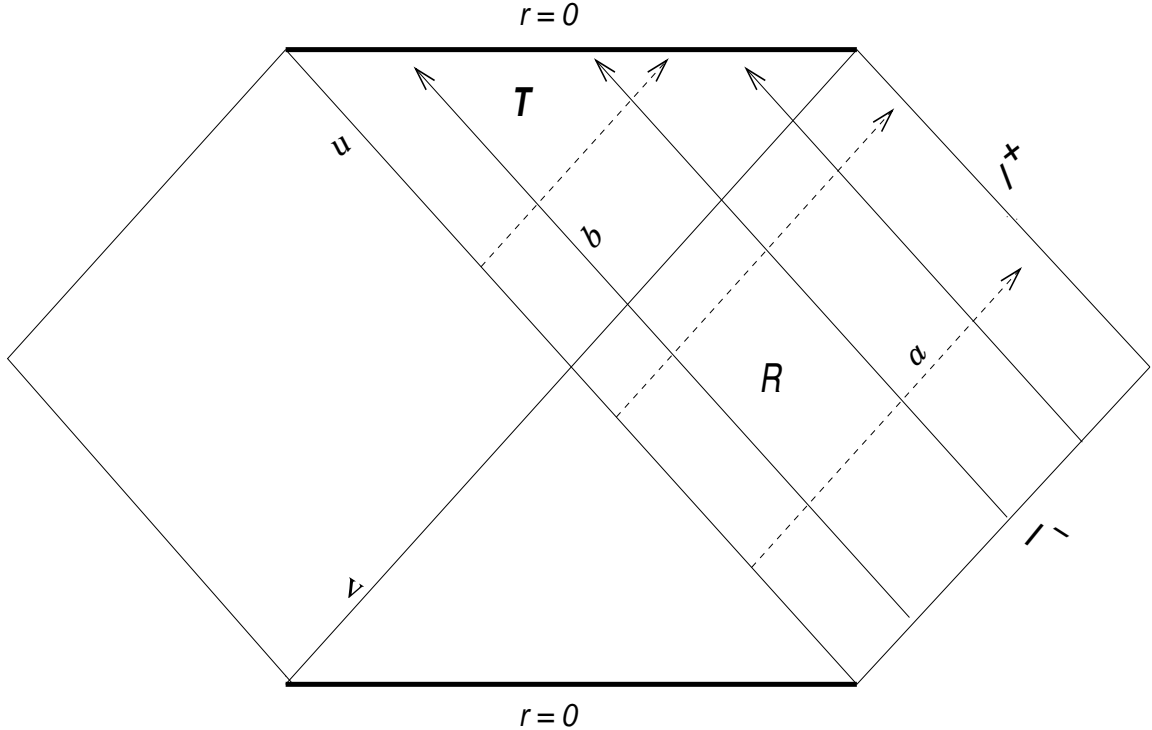,height=3.8in,width=6.0in}}
\label{F:painleve}
\caption{{\small The Penrose diagram of the Schwarzschild space-time. $u$ 
and $v$ are the lines $r = 2M$. $(R, T)$ region corresponds to the
line-element (\ref{painm1}). Lines $a$, $b$ are the light rays. $a$
are the outgoing light rays and these cannot be traced back past the
horizon $u$; $b$ are the ingoing light rays and these fall into the
singularity}}
\end{figure}
\noindent One can notice that the indefiniteness in the sign in coordinate 
transformation is same as encountered in Lemaitre coordinate. The line
elements in Eq.~(\ref{painfull}) correspond to the ingoing and
outgoing null geodesics. [Figure (2) shows the Penrose diagram of the
Schwarzschild space. As described earlier, $(r, \tau_P)$ for the upper
sign in the above line-element cover regions $R$ and $T$.] In terms of
$r$ and $\tau_P$, these are given by
\br
\!\!\!\!\!\!\!\!\!\!\!\!\!\!\!\!\!\!\!\!\!\!\!\!\!\!\!& &{\rm ingoing:}~ 
\tau_P + r - 2 {\sqrt{2Mr}} + 4M\log\l[\sqrt{r} + 
\sqrt{2M}\r]\,=\, v \,=\, 
{\rm constant} \nonumber \\
\!\!\!\!\!\!\!\!\!\!\!\!\!\!\!\!\!\!\!\!\!\!\!\!\!\!\!& & {\rm outgoing:}~ 
\tau_P - r - 2 {\sqrt{2Mr}} - 4M\log\l[\sqrt{r} + 
\sqrt{2M}\r]\,=\, u \, = \, 
{\rm constant} 
\er
\par
In the case of Lemaitre coordinate, we identified that the physical
extension of the external Schwarzschild space-time to the region
inside the Schwarzschild sphere has a doubled valued nature by
splitting the space-time defined using R- and T- regions. We make use
of the same formalism here to analyze the space-time structure of the
\pl metric.
\par
In our earlier discussion of R- and T- regions in 
Sec.~(\ref{subsec:rtregions}), these regions were defined for the 
line elements not involving mixed components $g_{0i}$. The mixed components, 
for spherically symmetric space-times, can be set equal to zero inasmuch as 
rotation is absent by the assumption of sphericity. Hence in the literature, 
the definition of R- and T- regions for space-times with the mixed components 
is not considered. In order to identify R- and T- regions in the \pl 
coordinate, we first obtain below a transformation connecting \pl and 
Lemaitre coordinates $(\tau, R, \theta, \phi)$. 
\par
Noting that the constant time slice is a flat Euclidean space for the \pl 
metric, we transform the coordinate variables of Lemaitre as

\b
\tau_P = \tau, ~~~~ r = (2M)^{1/3} \l(\frac{3}{2} (R - \tau) \r)^{2/3}
\label{maintrans}
\e
\noindent to go from Lemaitre to \pl line elements. Under these 
transformations, the Lemaitre line element in (\ref{lemaitremetric1}) now 
translates into \pl line element in (\ref{painm1}). The other \pl 
line element in (\ref{painm2}) is obtained by repeating the same steps by 
starting from the Lemaitre coordinate in (\ref{lemaitremetric2}), which is the 
time reversal of the line element in (\ref{lemaitremetric1}).  
\par
The criterion of R- and T- regions for \pl metric can be easily 
obtained from Eq.~(\ref{limcond}). For the line elements in (\ref{painm1}) and 
(\ref{painm2}), the inequality satisfied for the R- region is $r > 2M$. Hence, 
for the two line elements of the \pl metric, R- and T- regions are the 
same -- the whole of the space-time is doubly mapped. As mentioned in the 
earlier subsection, the contribution to the amplitude of absorption/emission 
will be from two mutually exclusive paths.     

\section{Semiclassical treatment}
\label{sec:semiclas}
In this section, we consider particle production in the two coordinate
systems described in the previous section by the semiclassical method
which was outlined in Sec.~(\ref{sec:hawking}). The semiclassical
wavefunction for the quantized scalar field and consequently the
semiclassical propagator is written down and by regularizing the
singularity in the action functional by complex extension and Hawking
radiation is obtained for both these coordinate systems. Before, we
proceed with the application of the method of complex paths in these
coordinates, we would like to point the readers that the method of
complex paths {\it does not} require the Killing vectors to be
time-like. Noting that the method requires to separate the non-linear
partial differential equation into an {\it ordinary} differential
equation for identifying the singularities, we need to use the ansatz,
\b
        S = - E \xi + f(\rho)
\label{eq:ansatz}
\e
where $\xi$ represents the coordinate associted to the Killing vector. 
However, the Killing vector {\it need not} coincide asymptotically with 
the Minkowski time. In the case of \pl (and Schwarzschild), as we will 
notice, $\xi$ coincides with the Minkowski time asymptotically while 
$\xi$ is a light cone coordinate in the case of Lemaitre and does not 
coincide with the Minkowski coordinate asymptotically.
 
\subsection{Lemaitre coordinate}
\label{subsec:lemaitres}
The method described in Sec.~(\ref{sec:hawking}) in obtaining the
thermal spectrum for the Schwarzschild coordinate will be used here
without modification. We consider the minimally, coupled, massless,
scalar field propagating in the Lemaitre background. The system we
first consider is that for in-going particles described by the metric
in Eq.~(\ref{lemaitremetric1}). Since the entire physics is contained
in the $(\tau, R)$ plane, we separate out the angular variables and
set $\Phi = \Psi(\tau, R) Y^m_l(\theta, \phi)$. The equation satisfied
by $\Psi$ is
\br
\label{lemsemi1}
& &\l[1-U^{2/3}\r] \l( \partial_{\ss U}^2 + \partial_{\ss V}^2 \r) \Psi - 
2 \l[1+U^{2/3}\r] \partial_{\ss U}\partial_{\ss V} \Psi \\ 
&+&\l[\frac{1}{U} - \frac{5U^{-1/3}}{3}\r] 
\l( \partial_{\ss U}^2 - \partial_{\ss V}^2 \r) \Psi
= \frac{16l(l+1)}{9U^{4/3}} \Psi 
\nonumber 
\er
where $U$ and $V$ are dimensionless parameters given by
\b
U = \frac{3}{4M}\l(R - \tau\r), \qquad V = \frac{3}{4M}\l(R + \tau\r)
\label{lemsemi2}.
\e
Note that the horizon is the surface $U=1$. Making the usual semiclassical 
ansatz for $\Psi$ as done in Eq.~(\ref{eqn:semi}) and expanding the $S$ in 
powers of $\hbar$ as done in Eq.~(\ref{eqn:exp}) one finds, to lowest order,
\b
\!\!\!\!\!\!\!\!\!\!\!\!\!\!\!\!\!\!\!\!\!\!\!\!\!\!\!\!\!\!\!\!\!\!\!\!\l(1-U^{2/3}\r)\l[ \l(\partial_{\ss U}S_0\r)^2 + \l(\partial_{\ss V}S_0\r)
^2 \r] 
- 2\l(1+U^{2/3}\r)\l(\partial_{\ss U}S_0\r)\l(\partial_{\ss V}S_0\r) 
+ \frac{16l(l+1)}{9U^{2/3}} = 0
\label{lemsemi3}.
\e
The above equation is just the Hamilton-Jacobi equation satisfied by a 
massless particle moving in the space time determined by the line element in  
(\ref{lemaitremetric1}). Specializing to the case $l = 0$ for simplicity, and 
making the ansatz $S_0 = -4M E V/3 + f(U)$, we obtain,
\b
\frac{df}{dU} = E\frac{1+U^{2/3} \pm 2U^{1/3}}{1-U^{2/3}}
\label{lemsemi4},
\e
where the $\pm$ signs arise from taking square roots. Notice that the
denominator is singular at $U=1$ only for the positive sign. This
singular solution for $f$ evidently corresponds to outgoing particles
[$(\pa S_0/\pa U) > 0$]. Therefore choosing the positive sign and
making the convenient change of variable $x^3 = U$, the solution to
Eq.~(\ref{lemsemi3}) is given by
\b
S_0[{\rm outgoing}] = -\frac{4ME}{3} V + 4ME\int\! dx \frac{x^2(1+x)}{1-x} 
\label{lemsemi5}.
\e 
[Note: The constant of integration $E$ is the ``energy" of the
particle corresponding to the dimensionless variable $V$. In order to
get the correct definition (of $E$ at the future infinity), we need to
set $S = - 4 M E V /3 + f(U)$.]  It is clear that the action function
is singular at the horizon $x=1$.  Following the method outlined in
Sec.~(\ref{sec:hawking}) and regularizing the singularity by complex
contour lying in the upper complex plane, we obtain 
\b 
S_0[{\rm emission}] = {\rm real \; part} + 8i\pi ME
\label{lemsemi7}
\e
In order to obtain the action for absorption of particles
corresponding to the ingoing particles [$(\pa S_0/\pa x) < 0$], we
have to repeat the above calculation using the metric given in
Eq.~(\ref{lemaitremetric2}). In this case, it is easy to see that the
only singular solution corresponds to in-going particles and so
\b
S_0[{\rm absorption}] = {\rm real \; part} -  8i\pi ME
\label{lemsemi8}
\e
where the minus sign arises from choosing the appropriate complex
contour given by the semiclassical prescription outlined in
Sec.~(\ref{sec:hawking}). [Note here that in the calculation of
$S_0$, we do not require the prescription of pushing the singularity at 
the horizon required in the case of Schwarzschild coordinate in 
Ref.~\cite{kt99}.] Constructing
the semiclassical propagator in the usual manner and taking the
modulus square we obtain the probability. From the earlier
discussions [see Sec.~(\ref{subsec:painleve})], we know that in calculating the
probability of absorption/emission there is an extra contribution due
to multiple mapping of part of the space-time and that the total
contribution to the probability is from four sets of the complex paths
satisfying the semiclassical ansatz. Taking this into account by
dividing $S_0$ by four, we obtain
\b
P[{\rm emission}] = \exp\left(-{8 \pi ME}\right)P[{\rm absorption}] 
\label{lemsemi9}.
\e
Thus, the exponential dependence of the energy allows us to give the thermal
interpretation to this result, which shows that the temperature of the
emission spectrum is the standard Hawking temperature. [Note that the 
if we translate the action in Eq. (\ref{lemsemi7}) in terms of the 
Schwarzschild coordinates (t,r), we find that 
$S_0[{\rm emission}]= -2 E t + F(r)$. The factor $2$ appears in
the action due to the multiple mapping of the part of the Lemaitre
coordinate (w. r. t. the Schwarzschild coordinate).] 
\par 
In Ref.\cite{kt99}, the authors have shown explicitly that close to the 
horizon, the terms containing the mass and angular part does not contribute 
significantly. Similar results hold in our case and the above analysis is 
therefore applicable to both massless and massive scalar particles.
\par
It has been shown by Davies\cite{davies76} that a freely falling detector will 
see a particle spectrum different from the thermal spectrum. Comparison with 
our result shows that there is no correspondence between the particles 
detected by the model detector and the particle spectrum obtained by the field 
theoretic analysis -- a result well known in other contexts as well. 
\par
In the next subsection, we will do similar analysis for the \pl metric 
and obtain the Hawking temperature by regularizing the singularity in the 
action functional by the complex extension. 

\subsection{\pl metric}
\label{subsec:painleves}
We proceed with the semiclassical analysis for the \pl metric, as done 
in the previous subsection. As before, we consider a minimally coupled 
massless scalar field $\Phi$. Since all the relevant physics is 
contained in the $(\tau_P,r)$ plane, we set $\Phi = \Psi(\tau_P, r) 
Y^m_l(\theta, \phi)$ beforehand. The equation satisfied by $\Psi$ is, 
\br
\label{painsemi1}
& &\frac{\pa^2\Psi}{\pa \tau_P^2}-2\sqrt{\frac{2M}{r}} \frac{\pa^2\Psi}{\pa 
\tau_P\pa r} -\frac{3}{2r}\sqrt{\frac{2M}{r}}\frac{\pa\Psi}{\pa \tau_P}
 \\
&-&\l(1-\frac{2M}{r}
\r)\frac{\pa^2\Psi}{\pa r^2} 
-\frac{2}{r}\l(1-\frac{M}{r}\r)\frac{\pa\Psi}{\pa r}-\frac{l(l+1)}
{r^2}\Psi = 0. \nonumber 
\er
Making the usual semiclassical ansatz for $\Psi$ as done in 
Eq.~(\ref{eqn:semi}) and expanding the $S$ in powers of $\hbar$ as done in 
Eq.~(\ref{eqn:exp}) one finds, to lowest order,
\b
\!\!\!\!\!\!\!\!\!\!\!\!\!\!\!\!\!\!\!\!\!\!\!\!\!\!\!\!\!\!\!\!\!\!-\l(\frac{\pa S_0}{\pa \tau_P}\r)^2 + 2\sqrt{2M \over r}
\l(\frac{\pa S_0}{\pa \tau_P} \r) \l(\frac{\pa S_0}{\pa r}\r) 
+ \l(1- {2M\over r}\r)\l(\frac{\pa S_0}{\pa r}\r)^2 + 
\frac{l(l+1)\hbar^2}{r^2} = 0.
\label{painsemi2}
\e
This is the Hamilton-Jacobi equation for a massless particle moving
in a space- time determined by the metric
Eq.~(\ref{painm1}). Introducing a dimensionless variable $\rho = r/2M$
and considering the s-waves, {\it i.e.} $l = 0$, for simplicity, the
solution to the above equation is easily found to be,
\b
S_0 = -E\tau_P + 2ME \int \! d\rho {\sqrt{\rho}(1\pm \sqrt{\rho}) \over \rho 
-1}.
\label{painsemi3}
\e
The denominator is singular at $\rho = 1$ only for the positive sign. Note 
that $\rho = 1$ corresponds to the black hole horizon. The singular solution
for $S_0$ evidently corresponds to outgoing particles [$(\pa S_0/\pa \rho) > 
0$]. Following the method outlined in Sec.~(\ref{sec:hawking}) and 
regularizing the singularity by a complex contour lying in the upper 
complex plane and using the arguments in Sec.~(\ref{subsec:painleve}), we 
obtain
\b
S_0[{\rm emission}] = {\rm real \; part} + 2i\pi ME.
\label{painsemi}
\e
In this case, the action that has been calculated is interpreted to be that 
for emission by analogy with that done in Sec.~(\ref{sec:hawking}). In 
order to obtain the action for absorption of particles corresponding to the
ingoing particles [$(\pa S_0/\pa \rho) < 0$], we have to consider the 
Hamilton-Jacobi equation for the metric in Eq.~(\ref{painm2}) and repeating 
the above calculations and using the arguments above we obtain 
\b
S_0[{\rm absorption}] = {\rm real \; part} -  2i\pi ME,
\label{painsemi5}
\e
where the minus sign arises from choosing the complex contour lying in the 
upper complex plane given by the semiclassical prescription outlined 
in Sec.~(\ref{sec:hawking}). Here again we notice that the calculation of 
$S_0$ does not require any prescription, just like in the Lemaitre coordinate,
as against the Schwarzschild coordinate. By constructing the semiclassical 
propagator in the usual manner and taking the modulus square to get the 
probability, we obtain 
\b
P[{\rm emission}] = \exp\left(-{8 \pi ME}\right)P[{\rm absorption}]. 
\label{painsemi6}
\e
Here again, the exponential dependence of the energy allows us to give the 
thermal interpretation to this result, which shows that the temperature of the
emission spectrum is the standard Hawking temperature. It may be noted that 
in our interpretation, we consider the amplitude for pair creation
both inside and outside the horizon. This is different from the treatment of 
Hawking radiation obtained recently by Parikh and Wilczek\cite{parikh}.
The authors considered Hawking radiation as a pair creation outside the
horizon, with the negative energy particle tunneling into the black hole. The
tunneling of particles produced just inside the horizon also contributes to
the Hawking radiation and is incorporated in our formalism.
 
\section{Conclusions}
\label{sec:conclusions}
	We have studied the spectrum of created particles in two
coordinate representations, \pl and Lemaitre, of the Schwarzschild
space-time for a linear, massless scalar field using the method of
complex paths. We have found that in both these coordinate
representations the spectrum of radiation is thermal and the
temperature is same as the standard Hawking temperature.  

The classical space-time analysis of these two coordinates is done
using R- and T- regions introduced by Novikov \cite{novikov64}. In
Ref.~\cite{hawking76}, Hartle and Hawking calculated the probability
amplitude for the emission and absorption by identifying the
particular path for the particles to cross the horizon, {\it i.e.} the
in-going and out-going paths which requires a detailed study of
particle trajectories.  As we have seen, analysis of these two
coordinates, \pl and Lemaitre, using R- and T- regions provides an
elegant method in understanding the global structure of the
space-time, thus making the detailed analysis of particle trajectories
unnecessary.

The physical effect of Hawking radiation is related to the decrease in
the mass of the black hole; the mass being converted into energy, thus
implying the radiation emitted should be covariant. In order to prove
the covariance of Hawking radiation, we would like to have a general
proof for all the coordinate representations of the Schwarzschild
space-time. However, a general result showing the `general covariance
of Hawking radiation' is really a difficult task. Hence, we have
considered two coordinate systems whose properties are different from
that of the Schwarzschild metric {\it i.e.}  both are non-static and
do not possess singularity at the horizon. Using the method of complex
paths, we have shown that in both these coordinate representations,
\pl and Lemaitre, the spectrum of particle emitted is thermal and the
temperature is the standard Hawking temperature($1/(8\pi M)$) and
hence indicating covariance of the Hawking radiation as far as these
coordinates are concerned. Thus, we have shown that even-though the
action satisfies a generally covariant HJ equation, when one transforms
coordinates, the action integral develops poles and this can be
interpreted by using Novikov's $R$ and $T$ regions analysis of the
space-time manifold.

The analysis which we have performed, in our paper, considers the
amplitude for pair creation for both inside and outside the
horizon. Our treatment is different from the treatment of Hawking
radiation obtained by Parikh and Wilczek \cite{parikh}. In our
approach, the tunneling of particles just produced inside the horizon
(tunneling in the context of non-relativistic quantum mechanics)as well
as the pair creation just outside the horizon (over the barrier
reflection) contributes to the Hawking radiation.

The only analysis that has been performed in the literature is that of
the particle spectrum of a freely falling detector. It has been shown
by Davies\cite{davies76} that a freely falling detector will see a
particle spectrum different from the thermal spectrum. Comparison with
results of Lemaitre coordinate, which is natural to a freely falling
observer, shows that there is no correspondence between the particles
detected by the model detector and the particle spectrum obtained by
the field theoretic analysis -- a result known in contexts of flat
space-times (see Ref.~\cite{schrambo98} and references therein).

\ack
The authors would wish to thank Prof. M.A.H. MacCallum for providing
with the English translation of Novikov's work before it's publication
in GRG. S.S.  is being supported by the Council of Scientific and
Industrial Research, India.


\section*{References}
\begin{thebibliography}{20}
\bibitem{schrambo98}
L.~Sriramkumar and T.~Padmanabhan, {\sl Probes of the 
vacuum structure of quantum fields in classical backgrounds}, 
Int.\ J.\ Mod.\ Phys.\ D {\bf 11} 1 (2002); gr-qc/9903054.
\bibitem{hawking75}
S.~Hawking, Commun.\ Math.\ Phys. {\bf 43}, 199(1975).
\bibitem{hawking76}
J.B.~Hartle and S.W.~Hawking, Phys.\ Rev.\ D \ {\bf 13}, 2188 (1976).
\bibitem{kt99}
K.~Srinivasan and T.~Padmanabhan, {\sl Particle production and complex 
path analysis}, Phys.\ Rev.\  D \ {\bf 60}, 24007 (1999).
\bibitem{landau3}
L.~D.~Landau and E.~M.~Lifshitz, {\sl Quantum Mechanics (Non-relativistic 
Theory)}, Course of Theoretical Physics, Volume~3 \ 
(Pergamon Press, New York, 1975).
\bibitem{candelas}
P. Candelas, {\sl Vacuum polarization in Schwarzschild spacetime}, 
Phys.\ Rev.\ D \ {\bf 21}, 2185 (1980).
\bibitem{visser}
M. Visser, {\sl Gravitational vacuum polarization: I, II, III}, 
Phys.\ Rev.\ D \ {\bf 54}, 5103 - 5128 (1996).
\bibitem{landau2}
L.~D.~Landau and E.~M.~Lifshitz, {\sl Classical Theory of Fields}, Course of 
Theoretical Physics, Volume~2  \  (Pergamon Press, New York, 1975).
\bibitem{parikh}
M.~K.~Parikh and F.~Wilczek, {\sl Hawking Radiation as Tunneling}, 
Phys.\ Rev.\ Lett. \ {\bf 85}, 5042 (2000); hep-th/9907001.
\bibitem{ralf}
R. Schutzhold, {\sl On the Hawking effect}, Phys.\ Rev.\ D\ {\bf 64},
024029 (2001); gr-qc/0011047.
\bibitem{unruh81}
W.~G.~Unruh, Phys.\ Rev.\ Lett.\ {\bf 46}, 1351 (1981).
\bibitem{jacobson}  
T.~Jacobson, Phys.\ Rev.\  D\ {\bf 44}, 1731 (1991); W.~G.~Unruh, 
{\it ibid} \ {\bf 51}, 2827 (1995); S.~Corley and T.~Jacobson, {\it ibid} 
\ {\bf 59}, 124011 (1999).
\bibitem{paddy91}
T.~Padmanabhan, Pramana--J.\ Phys.\  {\bf 37}, 179 (1991).
\bibitem{paddycqg}
T.~Padmanabhan, {\sl Why does an accelerated detector click?}, 
Class.\ Quan.\ Grav. {\bf 2}, 117 (1985).
\bibitem{bandd82}
N.~D.~Birrel and P.~C.~W.~Davies, {\sl Quantum Field in Curved Space} \ 
(Cambridge University Press, Cambridge, England, 1982). 
\bibitem{novikov64}
I.~D.~Novikov, Communications of the Shternberg state 
Astronomical Institute, {\bf 132}, 3-42 (1964).
\bibitem{kraus}
P.~Kraus and F.~Wilczek, {\sl Some Applications of a Simple 
Stationary Line Element for the Schwarzschild Geometry}, gr-qc/9406042.
\bibitem{davies76}
P.C.W.~Davies, Rep.\ Prog.\ Phys.\ {\bf 41}, 1315 (1976).
\bibitem{letaw}
J.~R.~Letaw, Phys.\ Rev.\ D. {\bf 23}, 1709 (1981); J.~R.~Letaw and
J.~D.~Pfautsch, {\it ibid} {\bf 24}, 1491 (1981).
\end{thebibliography}
\end{document}